\documentclass[useAMS,usenatbib]{mn2e}

\usepackage{natbib}
\usepackage{graphicx}
\usepackage{amssymb}
\usepackage{color}
\usepackage[dvipsnames]{xcolor}
\usepackage{caption}
\usepackage{lipsum,graphicx,multicol}
\usepackage{float}
\usepackage[fleqn]{amsmath}
\usepackage{subcaption}

\title[GW mergers of accreting BBHs in AGN discs]{Gravitational wave mergers of accreting binary black holes in AGN discs} 
\author[ ]
{W. Ishibashi$^{1,2,3}$\thanks{E-mail: wako.ishibashi@physik.uzh.ch} and M. Gr{\"o}bner$^{4}$ 
\footnotemark[0]\\
\footnotemark[0]\\
$^{1}$Physik-Institut, Universit{\"a}t Z{\"u}rich, Winterthurerstrasse 190, 8057 Z{\"u}rich, Switzerland \\
$^{2}$Istituto Ricerche Solari (IRSOL), Universit$\grave{a}$ della Svizzera italiana (USI), 6605 Locarno Monti, Switzerland \\
$^{3}$Euler Institute, Universit$\grave{a}$ della Svizzera italiana (USI), 6900 Lugano, Switzerland \\ 
$^{4}$Department of Physics, ETH Z{\"u}rich, Otto-Stern-Weg 1, Z{\"u}rich, Switzerland
}

\begin{document}

\pdfminorversion=4

\date{ Accepted 2024 February 19. Received 2024 February 14; in original form 2023 March 27 }

\pagerange{\pageref{firstpage}--\pageref{lastpage}} \pubyear{2012}

\maketitle

\label{firstpage}

\begin{abstract} 
Binary black hole (BBH) evolution in the discs of active galactic nuclei (AGN) is a promising channel for gravitational wave (GW)-driven mergers. It is however unclear whether binaries interacting with the surrounding disc undergo orbital contraction or expansion. We develop a simple analytic model of accreting BBHs in AGN discs to follow the orbital evolution from the disc-dominated regime at large separations into the GW-driven regime at small separations (the coupled `disc+GW'-driven evolution). We obtain that accreting binaries expand in thick discs with aspect ratio greater than a critical value ($> h_\mathrm{crit}$); whereas accreting binaries contract and eventually merge in thin discs ($< h_\mathrm{crit}$). Interestingly, accreting BBHs can experience faster mergers compared to non-accreting counterparts, with a non-monotonic dependence on the disc aspect ratio. The orbital contraction is usually coupled with eccentricity growth in the disc-dominated regime, which lead to accelerated inspirals in the GW-driven regime. We quantify the resulting BBH merger timescales in AGN discs ($\tau_\mathrm{merger} \sim 10^5 - 10^7$ yr) and estimate the associated GW merger rates ($\mathcal{R} \sim (0.2 - 5) \, \text{Gpc}^{-3} \text{yr}^{-1}$). Overall, accreting binaries may efficiently contract and merge in thin discs, hence this particular BBH-in-AGN channel may provide a non-negligible contribution to the observed GW merger event rate. 
\end{abstract}  



\section{Introduction}

Gravitational wave (GW) signals have been routinely detected by Advanced LIGO and Advanced Virgo in the last observing runs. The third gravitational wave transient catalog (GWTC-3) includes 90 GW events, with the great majority originating from binary black hole (BBH) mergers \citep{LVK_GWTC_3}. Despite the increasing number of GW detections, the astrophysical origin of BBH mergers is still much debated. A number of different binary formation channels have been discussed in the literature -- from isolated evolution in galactic fields to dynamical formation in dense star clusters -- with multiple sub-channels and mixtures thereof \citep[e.g.][and references therein]{Zevin_et_2021, Belczynski_et_2022}. 

An alternative binary formation path that is currently gaining interest in the community is the evolution of BBHs in the accretion discs of active galactic nuclei (AGN) \citep{Stone_et_2017, Bartos_et_2017, McKernan_et_2018, Tagawa_et_2020, Samsing_et_2022}. In such a scenario, gaseous interactions in the AGN disc assist the binary orbital decay, eventually leading to GW-driven mergers. This particular BBH evolution channel has a number of interesting physical implications: the potential occurrence of electromagnetic counterparts \citep{Ford_et_2019, Perna_et_2019}, the possible growth of intermediate-mass black holes (IMBH) \citep{McKernan_et_2012}, and the presence of eccentric binaries and retrograde orbiters within AGN discs \citep{Samsing_et_2022, Secunda_et_2021}. There are actually some indications of eccentric BBH mergers in GWTC-3 \citep{Romero-Shaw_et_2022} and a non-zero orbital eccentricity (e.g. GW190521) may already be suggestive of an AGN disc origin. 

It is generally assumed that BBH mergers are facilitated in the gaseous environment of AGN accretion discs.   
Early theoretical studies suggest that the orbital angular momentum is efficiently extracted by an external disc \citep{Pringle_1991}, and that the semi-major axis decreases while the eccentricity increases, as shown in the first smoothed particle hydrodynamic simulations \citep{Artymowicz_et_1991}. The semi-major axis decay is coupled with the orbital eccentricity growth, with a circumbinary gap possibly forming as a result of the disc-binary interactions \citep[][and references therein]{Artymowicz_Lubow_1994, Lubow_Artymowicz_2000}. In this picture, the binary loses angular momentum to the surrounding disc, leading to orbital contraction. Such gas-assisted orbital decay may eventually lead to GW-driven coalescence \citep{Armitage_Natarajan_2005, Hayasaki_2009}.  

However, recent numerical simulations indicate that accreting binaries gain net angular momentum from the circumbinary disc, resulting in orbital expansion rather than contraction \citep{Munoz_et_2019, Moody_et_2019, Duffell_et_2020}. If this is always the case, accreting BBHs in AGN discs may never reach the GW-driven merger stage. In contrast, other studies suggest that the direction of the binary orbital evolution -- either expansion or contraction -- depends on the underlying parameters, such as the disc aspect ratio \citep{Tiede_et_2020, Heath_Nixon_2020}. The actual outcome is now much debated: what determines whether accreting binaries contract or expand, and under which circumstances are BBH mergers accelerated or hindered in AGN discs? The question of binary contraction versus expansion has also important implications on the BBH merger timescales and ultimately on the GW event rates. 

In this paper, we develop an idealised analytic model of accreting BBHs in AGN discs. We compute the binary orbital evolution from the disc-driven regime on large scales (dominated by disc-binary interaction) into the GW-driven regime on small scales (dominated by GW emission). Presuming that accreting binaries can contract in AGN discs, we wish to quantify the resulting BBH merger timescales and associated GW event rates. We previously considered the simple case of non-accreting BBHs \citep{Ishibashi_Groebner_2020} and estimated the corresponding GW merger rate densities \citep{Groebner_et_2020}. Here we investigate a more general form of the binary orbital evolution in AGN discs by explicitly including mass accretion. 

The paper is structured as follows. In Section \ref{Section_model}, we introduce our toy model of accreting BBHs evolving in AGN discs, and derive the coupled `disc+GW'-driven evolution equations (Section \ref{Section_coupled_evolution}). We analyse the resulting BBH orbital evolution to determine when accreting binaries expand or contract in AGN discs (Section \ref{Section_contraction_expansion}). We quantify the associated BBH merger timescales in Section \ref{Section_merger_timescale}, and estimate the corresponding GW merger rates in Section \ref{Section_GW_rates}. We compare our analytic model results with some recent numerical simulations, and discuss the global implications of BBH mergers in AGN discs as an alternative channel for GW sources (Section \ref{Section_discussion}).  


\section{Analytic model of accreting binary black holes in AGN discs}
\label{Section_model}

Black hole binaries in AGN discs may arise from a variety of astrophysical origins, from capture objects in nuclear star clusters to in-situ star formation in self-gravitating accretion discs. Close encounters between stellar-mass objects orbiting in the AGN disc may facilitate the formation of stably bound eccentric binaries, possibly via some form of gas-assisted binary capture \citep{Rowan_et_2023, DeLaurentiis_et_2023, Whitehead_et_2023}.

We consider a BBH system of mass $M_b$ in an AGN accretion disc around a central super-massive black hole (SMBH) of mass $M_\mathrm{SMBH}$. We assume a geometrically thin disc (with disc aspect ratio $h = H/r \ll 1$) following Keplerian rotation $\Omega(r) = \sqrt{\frac{G M_\mathrm{SMBH}}{r^3}}$. The radial distance $r$ in the AGN disc can be parametrized in units of the Schwarzschild radius, as $r = \zeta R_S  = \zeta \frac{2 G M_\mathrm{SMBH}}{c^2}$. The binary stability criterion requires that the semi-major axis must be much smaller than the Hill radius, i.e. $a \ll R_H = r \left( M_b/3 M_\mathrm{SMBH}\right)^{1/3}$. Following energetic encounters in the AGN gaseous disc, the resulting binaries may become tightly bound, with the semi-major axis being only a small fraction of the Hill radius.

The binary interacts with the surrounding disc via orbital resonances, such as Lindblad and corotation resonances, which mediate the transfer of angular momentum between the two \citep{Lynden-Bell_Kalnajs_1972, Goldreich_Tremaine_1980}. A gap or cavity may form in the circumbinary disc if the resonant torque -- that tends to open the gap -- exceeds the viscous torque, which tends to fill the gap \citep[][and references therein]{Lubow_Artymowicz_2000}. In our previous study \citep{Ishibashi_Groebner_2020}, we simply assumed that the BBH resides in an empty cavity and neglected any mass transfer between the disc and the binary.  

In more realistic situations, mass accretion can occur from the disc to the binary, mainly in the form of narrow accretion streams penetrating within the cavity and leading to the formation of mini-discs around each black hole, as seen in several numerical simulations \citep{MacFadyen_Milosavljevic_2008, Shi_et_2012, Farris_et_2014, Tang_et_2017, Munoz_et_2019, Moody_et_2019}. Since the accreting matter advects energy and angular momentum, it can have a major impact on the orbital evolution of the BBH. While the binary loses angular momentum through gravitational resonant torques, it can gain angular momentum through accretion torques. 
Here we develop a simple analytic model for the orbital evolution of accreting BBHs based on the torque balance and subsequent GW emission. 


\subsection{Evolution of binary orbital elements}

We consider a BBH of total binary mass $M_b = M_1 + M_2$ on an elliptic orbit with semi-major axis $a$ and orbital eccentricity $e$. In the following, we work in the reference frame of the binary system. The total energy of the binary is
\begin{equation}
E_b = - \frac{G M_b \mu}{2a} , 
\label{Eq_energy}
\end{equation}
where $\mu = \frac{M_1 M_2}{M_b} = \frac{q}{(1+q)^2} M_b$ is the reduced mass and $q = M_2/M_1$ is the binary mass ratio. 
We obtain by differentiating equation \ref{Eq_energy} with respect to time 
\begin{equation}
\frac{\dot{E}_b}{E_b} = - \frac{\dot{a}}{a} + \frac{\dot{M}_b}{M_b} + \frac{\dot{\mu}}{\mu} . 
\end{equation}
The orbital angular momentum of the binary is given by
\begin{equation}
L_b = \mu a^2 \Omega_b \sqrt{1-e^2} , 
\label{Eq_angmom}
\end{equation}
where $\Omega_b = \sqrt{\frac{G M_b}{a^3}}$ is the binary orbital frequency. 
The time derivative of equation \ref{Eq_angmom} gives
\begin{equation}
\frac{\dot{L}_b}{L_b} = \frac{\dot{a}}{2a} - \frac{\dot{e} e}{1-e^2} + \frac{\dot{M}_b}{{ 2 M_b}} + \frac{\dot{\mu}}{\mu} . 
\end{equation}
From equations \ref{Eq_energy} and \ref{Eq_angmom}, the binary energy and angular momentum are related by 
\begin{equation}
E_b = - \frac{\Omega_b L_b}{2 \sqrt{1-e^2}} . 
\end{equation} 

The binary mainly interacts with the surrounding circumbinary disc through orbital resonances. The orbital elements evolve as a result of complex disc-binary interactions, and a number of idealised assumptions must be adopted in our analytic modelling. We assume that the disc-binary interaction can be approximated as an adiabatic process and that the non-axisymmetric potential perturbations are small around the average binary potential. 
In this framework, the binary energy dissipation rate can be related to the rate of change of the angular momentum via a characteristic frequency of the system:
\begin{equation}
\dot{E_b} = \Omega_p \dot{L_b} .
\label{Eq_energy_dissipation}
\end{equation}
As a characteristic frequency we take the pattern speed $\Omega_p = \frac{l}{m} \Omega_b$ (where $l$ and $m$ are integers) associated to the strongest resonance torque (see Sect. \ref{Sect_resonances}). For simplicity, we do not consider eccentric discs and assume that the cavity maintains a quasi-circular shape throughout the binary orbital evolution, which is closely related to the assumption of small non-axisymmetric potential perturbations. An analogous approach is adopted in the analytic model of \citet{Hayasaki_2009} to compute the orbital evolution of eccentric binaries -- even with eccentricities approaching unity. 

By combining the above relations, we can write down the evolution equations of the semi-major axis and orbital eccentricity as
\begin{equation}
\frac{\dot{a}}{a} = 2 \frac{l}{m} \frac{\dot{L}_b}{\mu a^2 \Omega_b} + \frac{\dot{M}_b}{M_b} + \frac{\dot{\mu}}{\mu} 
\label{Eq_a}
\end{equation}  
\begin{equation}
\frac{\dot{e} e}{1-e^2} = \frac{\dot{L}_b}{\mu a^2 \Omega_b} \left( \frac{l}{m} - \frac{1}{\sqrt{1-e^2}} \right) + \frac{\dot{M}_b}{M_b} + \frac{3}{2} \frac{\dot{\mu}}{\mu} 
\label{Eq_e}
\end{equation}
In our previous work \citep{Ishibashi_Groebner_2020}, we ignored the accretion terms ($\dot{M}_b = \dot{\mu} = 0$).
Here we relax this assumption and consider a non-zero accretion rate $\dot{M}_b = \dot{M}_1 + \dot{M}_2$, where $\dot{M}_1$ and $\dot{M}_2$ are the individual accretion rates onto the primary and secondary BH, respectively. Introducing the relative accretion fraction $q_a = \dot{M}_2/\dot{M}_1$, the rate of change of the reduced mass can be written as
\begin{equation}
\frac{\dot{\mu}}{\mu}
= \frac{q^2 + q_a}{q(1+q_a)} \frac{\dot{M}_b}{M_b} , 
\end{equation} 
Numerical simulations indicate that the accretion rate onto the secondary is usually greater than that onto the primary \citep[][]{Lubow_Artymowicz_2000,Farris_et_2014}, and a simple fitting formula for the relative accretion fraction is provided by \citet{Duffell_et_2020}
\begin{equation}
q_a = \frac{1}{0.1 + 0.9q}.
\end{equation}


\subsection{Orbital resonances}
\label{Sect_resonances}

The binary interacts with the surrounding disc via orbital resonances. 
Following the linear theory of resonances \citep{Goldreich_Tremaine_1980, Artymowicz_Lubow_1994}, the binary potential can be decomposed into a sum of Fourier components:
\begin{equation}
\Phi(r, \theta, t) = \sum_{m,l} \phi_{m,l}(r) \exp [i \left( m \theta - l \Omega_b t \right)] . 
\end{equation}
Each potential harmonic $\phi_{m,l}$ rotates with pattern speed $\Omega_p = (l/m) \Omega_b$ (where $l$ is the time-harmonic number and $m$ is the azimuthal number), and each $\phi_{m,l}$ can give rise to a corotational resonance and two Lindblad resonances (inner and outer). The corotation resonance in the circumbinary disc is located at radius
\begin{equation}
r_\mathrm{CR} = \left( \frac{m}{l} \right)^{2/3} a , 
\end{equation}
while the Lindblad resonances are located at radii:
\begin{equation}
r_\mathrm{LR} = \left( \frac{m \pm 1}{l} \right)^{2/3} a , 
\end{equation}
with the upper (lower) sign corresponding to the outer (inner) Lindblad resonance. 

Analytic expressions for the torques exerted at the orbital resonances can be found in e.g. \citet{Goldreich_Tremaine_1979, Artymowicz_Lubow_1994}. The torque at corotation is given by 
\begin{equation}
T_\mathrm{CR, ml} = \frac{m \pi^2}{2} \left[ \frac{\phi_{ml}^2}{d\Omega_b/dr} \frac{d}{dr} \left( \frac{\Sigma}{B} \right) \right] , 
\end{equation}
where $\Sigma$ is the disc surface density at the resonance location, and $B(r) = \Omega(r) + \frac{r}{2} \frac{d\Omega}{dr}$ is the Oort parameter (equal to $\Omega/4$ for Keplerian rotation). The torques at the Lindblad resonances are given by
\begin{equation}
T_\mathrm{LR,ml} = - m \pi^2 \Sigma \frac{\vert \Psi_{ml} \vert^2}{r dD/dr} , 
\end{equation}
where $\Psi_{ml} = (\lambda \mp 2m) \phi_{ml}$ and $r dD/dr = \mp 3 \frac{l^2}{m \pm 1} \Omega_b^2$, again with the upper (lower) sign corresponding to the outer (inner) Lindblad resonance. Hence the Lindblad torques are 
\begin{equation}
T_\mathrm{LR,ml} = - m \pi^2 \Sigma \frac{(\lambda \mp 2m)^2 \phi_{ml}^2 (m \pm 1)}{\mp 3 l^2 \Omega_b^2} , 
\label{Eq_T_LR}
\end{equation}
with $\lambda = - (m+1)$ and $\lambda = m$ for the outer and inner Lindblad resonance, respectively; while the corotation torque is 
\begin{equation}
T_\mathrm{CR, ml} = - \frac{4}{3} m \pi^2 \frac{\phi_{ml}^2 \Sigma}{\Omega_b^2} , 
\label{Eq_T_CR}
\end{equation}
for a disc surface density following a power-law radial profile (e.g. $\Sigma \propto r^{-p}$ with $p = 1/2$). 

A gap or cavity may be cleared in the disc when the resonant torques exceed the disc viscous torque  \citep{Lubow_Artymowicz_2000}. According to the linear theory, the magnitude of the potential harmonic scales with eccentricity as $\phi_{m,l} \propto e ^{\vert m-l \vert}$, and thus the resonant torques scale as $T_{ml} \propto e^{2 \vert m - l \vert}$ \citep{Artymowicz_Lubow_1994}. As a consequence, the strongest torque applied to the disc is at the $m = l$ resonances, which are primarily responsible for truncating the disc. These are followed by higher order resonances ($m-l$ = 1, 2, ...), which can also lead to disc truncation and binary eccentricity growth. 

For typical binaries of comparable mass and low-to-moderate eccentricity, the dominant resonance is the $(m,l) = (2,1)$ outer Lindblad resonance, which governs the binary orbital evolution and also maintains the disc edge in equilibrium \citep{Artymowicz_et_1991, Lubow_Artymowicz_2000}. Indeed, the inner edge of the disc is determined by the outer Lindblad resonance, i.e. the location of the lowest order resonance for which the resonant torque equals the viscous torque \citep{Lubow_Artymowicz_2000, Hayasaki_2009}. Numerical simulations suggest that the cavity extends to about twice the semi-major axis, with the radial extent increasing with increasing eccentricity \citep{Artymowicz_Lubow_1994, Moesta_et_2019}.  We thus assume the inner edge of the circumbinary disc to be located at $r_{in} = 2a (1+e)$. 

Mass flows can develop and take place through the gap, mainly in the form of gas streams that feed the central binary. Although such accretion streams penetrate the cavity, the mass transfer occurs through an almost empty gap \citep{Artymowicz_Lubow_1996}. In fact, the narrow gas stream only fills a small fraction of the gap area such that the surface density is much lower than at the inner edge of the disc. The cavity may be significantly depleted, with typical gas depletion by two orders of magnitude in surface density, as also found in recent 3D smoothed particle hydrodynamics simulations \citep{Hirsh_et_2020}. Moreover, the mass accretion rate can be strongly suppressed in thin discs, with only a reduced fraction of the available gas flowing across the cavity \citep{Ragusa_et_2016}. 

Several resonances may operate in the circumbinary disc region. To estimate the relative importance of the resonant torques, we numerically compute the ratios of the Lindblad (inner and outer) and corotation resonances from equations \ref{Eq_T_LR}-\ref{Eq_T_CR}. For the dominant $(m,l) = (2,1)$ resonances, we obtain 
\begin{equation}
\frac{T_\mathrm{OLR}}{T_\mathrm{CR}} \sim - \frac{147}{4} \frac{\Sigma_\mathrm{OLR}}{\Sigma_\mathrm{CR}} , 
\label{Eq_ratio_OLR_CR}
\end{equation}
and 
\begin{equation}
\frac{T_\mathrm{OLR}}{T_\mathrm{ILR}} \sim - \frac{49}{12} \frac{\Sigma_\mathrm{OLR}}{\Sigma_\mathrm{ILR}} , 
\label{Eq_ratio_OLR_ILR}
\end{equation}
where the surface densities are evaluated at the corresponding resonance locations. 
Since the corotation and inner Lindblad resonances are located inside the depleted cavity region ($r_\mathrm{CR} = 2^{2/3} a \cong 1.6 a$, $r_\mathrm{ILR} = a$), their surface densities must presumably be smaller than that at the outer Lindblad resonance, which is located very close to the inner edge of the disc ($r_\mathrm{OLR} = 3^{2/3} a \cong 2.1 a$). The torque due to the outer Lindblad resonance should thus dominate over the corotation and inner Lindblad resonance torques. In fact, the ratios of the resonant torques are greater than unity ($T_\mathrm{OLR}/T_\mathrm{CR} \gg 1$, $T_\mathrm{OLR}/T_\mathrm{ILR} \gg 1$), even for comparable surface densities. 
Looking at the $(m,m-1)$ resonances (whose torque scale in the same way as the $(2,1)$ resonance torque), we find that as $m$ gets larger $T_\mathrm{OLR}$ dominates $T_\mathrm{CR}$ even more than in the $(2,1)$ case, allowing us to disregard $T_\mathrm{CR}$ in this case as well. With increasing $m$, $T_\mathrm{OLR}$ and $T_\mathrm{ILR}$ become of comparable opposite strength, leading to cancellation.

So overall only the $(m,m-1)$ outer Lindblad resonances with small values of $m$ persist, justifying our focus on the $(2,1)$ outer Lindblad resonance at the outermost position. As a consequence, the total gravitational resonant torque may be roughly approximated as: $T_\mathrm{grav} \approx  T_\mathrm{OLR} + T_\mathrm{ILR} + T_\mathrm{CR} \sim T_\mathrm{OLR}$ \citep[see also][]{Hayasaki_2009}. 
 
In general, the outer Lindblad resonance increases the binary eccentricity, whereas the corotation and inner Lindblad resonances tend to damp the binary eccentricity. Given that the outer Lindblad resonance provides the dominant torque, the eccentricity growth should prevail over eccentricity damping. Furthermore, corotation resonances may be saturated in the gaseous disc, leading to reduced corotation torques, and such saturation may also favour eccentricity excitation \citep{Ogilvie_Lubow_2003, Goldreich_Sari_2003}. In the following, we assume that the binary orbital evolution is mainly governed by the dominant $(m,l) = (2,1)$ outer Lindblad resonance, which leads to eccentricity growth (the case of reduced eccentricity growth is briefly discussed in Sect. \ref{Section_contraction_expansion}). 

A cautionary note should be added here, with a few caveats to keep in mind. We adopt a simplified picture of disc orbital resonances and surface density distribution (e.g. in estimating the ratios of the resonant torques in equations \ref{Eq_ratio_OLR_CR}-\ref{Eq_ratio_OLR_ILR}). We neglect the complexities arising from a more realistic treatment of the cavity profile and associated high-order resonances, as well as the detailed disc structure and gas thermodynamics. Some aspects of this intricate problem have been addressed in recent numerical studies of disc-binary interaction. 

Hydrodynamic simulations suggest that gas can flow across the depleted gap, keeping a quasi-steady state disc structure \citep[e.g.][and references therein]{Chen_et_2020}. The corresponding radial profile may then modify the location of the orbital resonances within the disc. Corotation and high-order Lindblad resonances lying in the gap region will scale with the reduced density in the depleted cavity, while low-order Lindblad resonances located just beyond the gap edge will likely dominate. In a perturbed disc with a deep gap, the asymmetry of the OLR and ILR can be significantly altered and low-order resonances tend to move closer. In this context, the disc surface density profile --and in particular its power-law slope-- can have a major effect in setting the overall torque direction and magnitude. For instance, in the case of a steep surface density gradient, the shifted higher-order resonances (e.g. $m = 3$) may cause the ILR torque to dominate over the OLR torque \citep{Chen_et_2020}. 

Furthermore, the thermodynamics of the gas in the circumbinary disc and mini-discs may play an important role.
2D hydrodynamical simulations taking into account the effect of BH feedback --via enhanced temperature profiles-- indicate that if the mini-discs are hotter than the background disc, the binary orbital evolution may turn from expansion to contraction \citep{Li_et_2022}. Refined 2D viscous hydrodynamic simulations of CBD accretion show that the subtle balance between viscous heating and radiative cooling --modelled via a simplified $\beta$ dynamical cooling prescription-- can substantially impact the binary orbital evolution \citep{Wang_et_2023}. Future numerical studies, possibly 3D simulations including magnetic effects, should help us better understand the complex dynamics of disc-binary interaction. 


\subsection{Torque balance}

In the disc-binary interaction process, the BBH can lose and/or gain angular momentum via gravitational and accretion torques. In the absence of mass accretion, the binary transfers angular momentum to the disc through gravitational torques ($T_{grav}$). When mass accretion is taken into account, angular momentum can be transferred from the disc to the binary by accretion torques ($T_{acc}$). The angular momentum balance is then given by 
\begin{equation}
\dot{L}_b = T_{grav} + T_{acc} \, .
\label{Eq_torque_balance}
\end{equation}

The inner edge of the disc is set by the competition between the gravitational resonant torque and the disc viscous torque. In an equilibrium situation, the gravitational torque --dominated by the outer Lindblad resonance-- is balanced by the viscous torque  evaluated at the inner edge of the disc, i.e. $T_{grav}  \sim T_\mathrm{OLR} = - T_{visc}$. Therefore the torque balance (equation \ref{Eq_torque_balance}) can be expressed as $\dot{L}_b = - T_{visc} + T_{acc}$.  

Following the classical $\alpha$-prescription in accretion disc theory \citep{Shakura_Sunyaev_1973}, the viscous torque is given by $T_{visc}(r) = -3 \pi \alpha c_s^2(r) \Sigma(r) r^2$, where $\alpha$ is the dimensionless viscosity parameter, $c_s = H \Omega$ is the local sound speed, and $\Sigma = \dot{M}/3 \pi \alpha c_s H \propto r^{-1/2}$ is the gas surface density, with $H$ being the disc scale height and $\dot{M}$ the mass accretion rate.  
Thus the viscous torque at radius $r$ in the disc is $T_{visc}(r) = \Omega \dot{M} r^2$. Assuming that the inner edge of the circumbinary disc is located at $r_{in} = 2a (1+e)$, the viscous torque at the disc inner edge is given by 
\begin{equation}
T_{visc}(r_{in}) 
= 4 a^2 (1+e)^2 \Omega \dot{M} . 
\label{Eq_T_visc}
\end{equation} 

Accretion flows can develop within the cavity, with matter accreting onto the primary and secondary BHs, and leading to the formation of a mini-disc around each BH. Accordingly we may decompose the induced change of angular momentum as $T_{acc} = \dot{L}_1 + \dot{L}_2$, where $\dot{L}_1$ and $\dot{L}_2$ are the contribution to $T_{acc}$ of the primary and secondary BH, respectively. We assume $\dot{L}_i$ to be proportional to the individual accretion rate $\dot{M}_i$, inversely proportional to the orbital frequency of the binary, and be a function of the mass $M_i$ and of some characteristic radius $r_i$ of the mini-disc. From dimensional analysis, each individual torque is then parametrized as 
\begin{equation}
\dot{L}_i = \frac{P_b}{\tau_{\nu,i}} \dot{M}_i \sqrt{G M_i r_i} , 
\label{Eq_dotL_i}
\end{equation}
where $P_b = 2 \pi /\Omega_b$ is the binary orbital period and $\tau_{\nu,i} = \frac{r_i^2}{\alpha_i c_{s,i} H_i}$ is the viscous timescale of the mini-disc.
As a consequence, the total torque due to accretion is
\begin{equation}
T_{acc} =  \dot{L}_1 + \dot{L}_2 = \frac{2 \pi}{\Omega_b} \frac{\dot{M}_b}{1+q_a} \left[ \alpha_1 c_{s,1}^2 + q_a \alpha_2 c_{s,2}^2 \right] ,
\end{equation} 
where $c_{s,i} = H_i \Omega_i = h_i \sqrt{\frac{G M_i}{r_i}}$ is the local sound speed in the mini-disc, and \textbf{$h_i = H_i/r_i$} is the disc aspect ratio of the mini-disc. For the characteristic radii $r_{1,2}$ we employ the tidal truncation radii $r_1 = 0.3 q^{- 0.3} a (1-e)$ and $r_2 = 0.3 q^{0.3} a (1-e)$ from \cite{Roedig_et_2014}. 
The final expression of the accretion torque is  
\begin{equation}
T_{acc} =  \frac{2 \pi}{\Omega_b} \frac{\dot{M}_b}{1+q_a} \left[ \frac{G M_b}{1+q} \frac{1}{0.3 a (1-e)} \left( \alpha_1 h_1^2 q^{0.3} + \alpha_2 h_2^2 q^{0.7} q_a \right) \right].
\label{Eq_T_acc}
\end{equation}

We note that the above parametrization of the torque due to accretion is similar to the one employed in \citet{Hayasaki_2009}.  In that work, the accretion torque was included in the torque balance, but $\dot{M}_b$ and $\dot{\mu}$ were neglected in their equations describing the energy dissipation rate and change rate of the orbital angular momentum. In contrast, here we self-consistently incorporate the accretion terms in both the torque balance (equation \ref{Eq_torque_balance}) and the binary orbital evolution equations (equations \ref{Eq_a}-\ref{Eq_e}). 


\subsection{Accretion rate estimates}
\label{Sect_accretion_rate_estimates}

In order to compute the accretion torque, we need an estimate of the accretion rate onto the binary system. If one just considers that a stellar-mass BBH accretes at some non-negligible fraction of the total mass accretion rate onto the central SMBH, the resulting binary accretion rate will be much larger than its Eddington rate (as $M_b \ll M_\mathrm{SMBH}$). In the absence of any feedback mechanism, an embedded stellar-mass BH would rapidly grow to become an intermediate mass black hole (IMBH) and deplete the surrounding gas in the AGN disc \citep{Tagawa_et_2022}. On the other hand, super-Eddington accretion will generate strong radiative and/or kinetic feedback on the surrounding environment, leading to the development of powerful bipolar outflows and/or jets \citep{Hu_et_2022_a, Tagawa_et_2022}. 

Due to the mass loss in the outflows, the gas inflow rate decreases with decreasing radius toward the centre, following a power law scaling of the form $\dot{M}(r) \propto r^p$ \citep[e.g.][and references therein]{Yuan_Narayan_2014}. As a consequence, the net accretion rate can be considerably reduced, with only a small fraction of the inflowing gas actually reaching the black hole. For instance, in the case of a convection-dominated accretion flow (CDAF), the inward (outward) angular momentum transport due to convection (viscosity) nearly cancel out, and the net accretion rate is indeed quite small \citep{Quataert_Gruzinov_2000}. 
More recent RHD simulations suggest that the mass inflow rate decays with a typical power-law index of $p \sim 0.5-0.7$ \citep{Hu_et_2022_a}. 

We therefore assume 
\begin{equation}
\dot{M} = \dot{M}_0 \times \left( \frac{r_{in}}{R_H} \right)^{p} , \quad 0 \leq p \leq 1
\end{equation}
where $r_{in}$ is the inner radius of the CBD and $R_H = r \left( M_b/3 M_\mathrm{SMBH}\right)^{1/3}$ is the Hill radius. The global mass accretion rate can be parametrized in units of the Eddington rate as $\dot{M}_0 = \dot{m} \dot{M}_E = \dot{m} L_E/\epsilon c^2$, where $L_E = \frac{4 \pi G c m_p M_\mathrm{SMBH}}{\sigma_T}$ is the Eddington luminosity of the central SMBH, $\dot{m}$ is the Eddington fraction, and $\epsilon$ is the radiative efficiency. For a typical AGN in the radiatively efficient accretion regime, $\epsilon \sim 0.1$ and $\dot{m} \sim 0.1$. 

Assuming that the gas surface density in the circumbinary disc is comparable to that of the underlying AGN disc with the same viscosity parameter $\alpha$, the CBD accretion rate is roughly given by
\begin{equation}
\dot{M}_\mathrm{cbd} = \frac{h_\mathrm{cbd}^2}{h^2} \frac{\sqrt{G M_b r_\mathrm{cbd}}}{\sqrt{G M_\mathrm{SMBH} r}}  \dot{M}_0 \left( \frac{r_{in}}{R_H} \right)^{p}, 
\label{Mdot_cbd_rev}
\end{equation} 
where $h_\mathrm{cbd}$ is the disc aspect ratio and $r_\mathrm{cbd}$ is some characteristic radius of the circumbinary disc. Given that the circumbinary accretion is typically set on the scale of the Hill radius, we adopt $r_\mathrm{cbd} \sim R_\mathrm{H}$ (cf. the so-called Hill accretion). 

Indeed an analogous result may be obtained by directly considering the order of magnitude scalings for the Hill accretion rate \citep{Rosenthal_et_2020, Choksi_et_2023, Li_et_2023}: 
\begin{equation}
\dot{M}_\mathrm{H} \sim \rho \times R_\mathrm{H} H \times v_\mathrm{H} 
\sim R_\mathrm{H}^2 \Omega \Sigma , 
\end{equation}
where $\rho \sim \Sigma/H$ is the disc density and $v_\mathrm{H} = \Omega R_\mathrm{H}$ is the Hill velocity.  
Assuming that the Hill radius and CBD scale height are comparable and neglecting numerical pre-factors we get 
\begin{equation}
\dot{M}_\mathrm{H}
\sim  H_\mathrm{cbd}^2 \Omega_\mathrm{cbd} \Sigma
\sim \frac{h_\mathrm{cbd}^2}{h^2}  \sqrt{ \frac{M_{b}}{M_\mathrm{SMBH}} \frac{r_\mathrm{cbd}}{r} } \dot{M}(r) , 
\end{equation}
which is basically equivalent to the CBD accretion rate (equation \ref{Mdot_cbd_rev}).

The strength of the feedback-driven outflow can be characterised by the power-law index $p$, and we consider three representative values: $p = 0$ (no outflow), $p = 1/2$ (fiducial case), and $p = 1$ (strong outflow). 
We note that in the fiducial case of moderate outflow with $p = 1/2$, the CBD accretion rate simply reduces to 
\begin{equation}
\dot{M}_\mathrm{cbd} 
=  \frac{h_\mathrm{cbd}^2}{h^2} \sqrt{ \frac{M_b}{M_\mathrm{SMBH}} \frac{r_{in}}{r} } \dot{M}_0 . 
\end{equation} 
Similarly, for the $p = 0$ case, the accretion rate is given by 
\begin{equation}
\dot{M}_\mathrm{cbd} = \frac{h_\mathrm{cbd}^2}{h^2} \frac{\sqrt{G M_b R_H}}{\sqrt{G M_\mathrm{SMBH} r}} \dot{M}_0 , 
\end{equation}
while for the $p = 1$ case, the corresponding accretion rate is
\begin{equation}
\dot{M}_\mathrm{cbd} = \frac{h_\mathrm{cbd}^2}{h^2}  \frac{\sqrt{G M_b r_{in}^2}}{\sqrt{G M_\mathrm{SMBH} r R_H}} \dot{M}_0  .  
\end{equation} 

Introducing the BBH-to-SMBH mass ratio $q_b = M_b/M_\mathrm{SMBH}$, we see that the CBD accretion rate scales as $\dot{M}_\mathrm{cbd} \propto  q_b^{1/2}$ for $p = 1/2$. Likewise, the accretion rate scales as $\dot{M}_\mathrm{cbd} \propto  q_b^{1/3}$ for $p = 1$, and $\dot{M}_\mathrm{cbd} \propto  q_b^{2/3}$ for $p = 0$. The latter scaling in the no-feedback case may be paralleled to the $\propto q_{th}^{2/3}$ scaling quoted in 3D simulation work of \citet{Li_et_2023}; although in more realistic situations including the gap opening effect, the scaling relation becomes more complicated. 

Finally, in the case of thin discs, the actual accretion rate onto the binary $\dot{M}_b$ is likely lower than the $\dot{M}_\mathrm{cbd}$ value given above. Indeed, 3D SPH simulations suggest that the mass accretion rate is suppressed in thin discs, such that  $\dot{M}_b = 10 h_\mathrm{cbd} \dot{M}_\mathrm{cbd}$ for $h_\mathrm{cbd} \leq 0.1$ \citep{Ragusa_et_2016}. This may be compared with the Eddington accretion rate for the combined binary mass $\dot{M}_\mathrm{E,b} = L_\mathrm{E,b} / \epsilon c^2$, where $L_\mathrm{E,b} = \frac{4 \pi G c m_p}{\sigma_T} M_b$. The resulting accretion ratio for the fiducial case ($p = 1/2$) is 
\begin{equation}
\frac{\dot{M}_b}{\dot{M}_\mathrm{E,b}} = 10 h_\mathrm{cbd} \left(\frac{h_\mathrm{cbd}}{h}\right)^2  \sqrt{\frac{M_\mathrm{SMBH}}{M_b} \frac{r_{in}}{r}} \dot{m} ,
\end{equation}
which scales as $\propto \sqrt{r_{in}} \propto \sqrt{2a(1+e)}$. 
We numerically verify that the accreting BBH system is unlikely to exceed the Eddington limit for typical parameters.\footnote{For fiducial parameters, the numerical value of the $\dot{M}_b/\dot{M}_\mathrm{E,b}$ ratio is $\sim 0.05, 0.15, 0.47$ for $a_0 = 1, 10, 100$ AU and $e_0 = 0.1$.}


\section{Coupled `disc+GW'-driven evolution} 
\label{Section_coupled_evolution}

Summarising the formulae from the previous section, the binary orbital evolution equations in the disc-driven regime are \begin{align}
\frac{\dot{a}}{a} = - 2 \frac{l}{m} \frac{T_{visc}}{\mu a^2 \Omega_b} + 2 \frac{l}{m} \frac{T_{acc}}{\mu a^2 \Omega_b} + \frac{\dot{M}_b}{M_b} \left[  1 +  \frac{q^2 + q_a}{q(1+q_a)}  \right] , 
\label{Eq_a_disc}
\end{align} 
\begin{align}
\frac{\dot{e} e}{1-e^2}  = \left( \frac{T_{visc}}{\mu a^2 \Omega_b} - \frac{T_{acc}}{\mu a^2 \Omega_b} \right) \left( \frac{1}{\sqrt{1-e^2}} - \frac{l}{m} \right) \nonumber \\ 
+ \frac{\dot{M}_b}{M_b} \left[ 1 + \frac{3}{2} \frac{q^2 + q_a}{q(1+q_a)}  \right] , 
\label{Eq_e_disc}
\end{align} 
where $T_{visc}$ and $T_{acc}$ are given by equation \ref{Eq_T_visc} and equation \ref{Eq_T_acc}, respectively; 
with the different estimates of the binary accretion rates discussed in Sect. \ref{Sect_accretion_rate_estimates}. 

Provided that the orbital separation shrinks enough, the binary eventually transitions from the disc-dominated regime into the GW-driven regime. In the latter case, the binary orbital evolution is determined by the GW-driven evolution equations \citep{Peters_1964} 
\begin{equation}
\dot{a} = - \frac{64}{5} \frac{G^3}{c^5} \frac{\mu M_b^2}{a^3} \frac{1}{(1-e^2)^{7/2}} \left( 1 + \frac{73}{24} e^2 + \frac{37}{96} e^4 \right),
\label{Eq_a_GW}
\end{equation}
\begin{equation}
\dot{e} = -  \frac{304}{15} \frac{G^3}{c^5} \frac{\mu M_b^2}{a^4} \frac{e}{(1-e^2)^{5/2}} \left( 1 + \frac{121}{304} e^2 \right). 
\label{Eq_e_GW}
\end{equation} 
Due to the steep eccentricity dependence, the resulting GW-driven inspiral can be considerably accelerated for high eccentricities. 

To follow the BBH orbital evolution from large scales down to small scales, we combine the disc-driven evolution equations with the corresponding GW-driven equations. The resulting coupled `disc+GW'-driven evolution is described by
\begin{equation}
\dot{a} =  \dot{a}_\text{disc} + \dot{a}_\text{GW},
\label{Eq_a_coupled}
\end{equation}
\begin{equation}
\dot{e} =  \dot{e}_\text{disc} + \dot{e}_\text{GW}, 
\label{Eq_e_coupled}
\end{equation}
where $(\dot{a}_\text{disc}, \dot{e}_\text{disc})$ and $(\dot{a}_\text{GW}, \dot{e}_\text{GW})$ are given in equations \ref{Eq_a_disc}-\ref{Eq_e_disc} and equations \ref{Eq_a_GW}-\ref{Eq_e_GW}, respectively. 
This coupled system of differential equations can be numerically integrated to obtain the temporal evolution of the semi-major axis $a(t)$ and orbital eccentricity $e(t)$. 

The following values are assumed as fiducial parameters of the model (unless otherwise stated): for the binary system, $M_b = 50 M_{\odot}$, $q = 0.5$, $a_0 = 1$ AU, $e_0 = 0.1$; for the AGN disc, $\alpha = 0.1$, $h = 0.01$, $\zeta = 10^5$, $M_\textrm{SMBH} = 10^7 M_{\odot}$, p = 1/2; and for the individual mini-discs, $\alpha_1 = \alpha_2 = 0.1$, $h_1 = h_2 = 0.1$. 


\section{Orbital evolution of accreting BBH: contraction or expansion?}
\label{Section_contraction_expansion}


\subsection{ Fiducial case }

We now analyse the binary orbital evolution resulting from the coupled `disc+GW'-driven evolution (equations \ref{Eq_a_coupled}-\ref{Eq_e_coupled}). In the simple case of non-accreting BBHs, only the viscous torque is present in equation \ref{Eq_a_disc}, and $T_{visc}$ determines the orbital evolution in the disc-driven regime. The negative viscous term implies $\dot{a}/a < 0$, and thus a uniform decrease of the semi-major axis, leading to systematic orbital contraction ($a$-decay). At the same time, the orbital eccentricity increases in the disc-dominated regime ($e$-growth), implying an enhanced orbital decay in the subsequent GW-driven regime -- due to the steep dependence on eccentricity. 

While non-accreting binaries always undergo orbital contraction, accreting BBHs can either contract or expand as a result of the interaction with the surrounding disc. In the disc-dominated regime, we see from equation \ref{Eq_a_disc} that the binary either contracts ($\dot{a}/a < 0$) or expands ($\dot{a}/a > 0$) depending on the relative importance of the viscous and accretion terms. Variations in accretion power can be parametrised by variations in the circumbinary disc aspect ratio (as $T_\mathrm{acc} \propto \dot{M}_b \propto h_\mathrm{cbd}$). 

\begin{figure}
  \centering
 \includegraphics[width=0.45\textwidth]{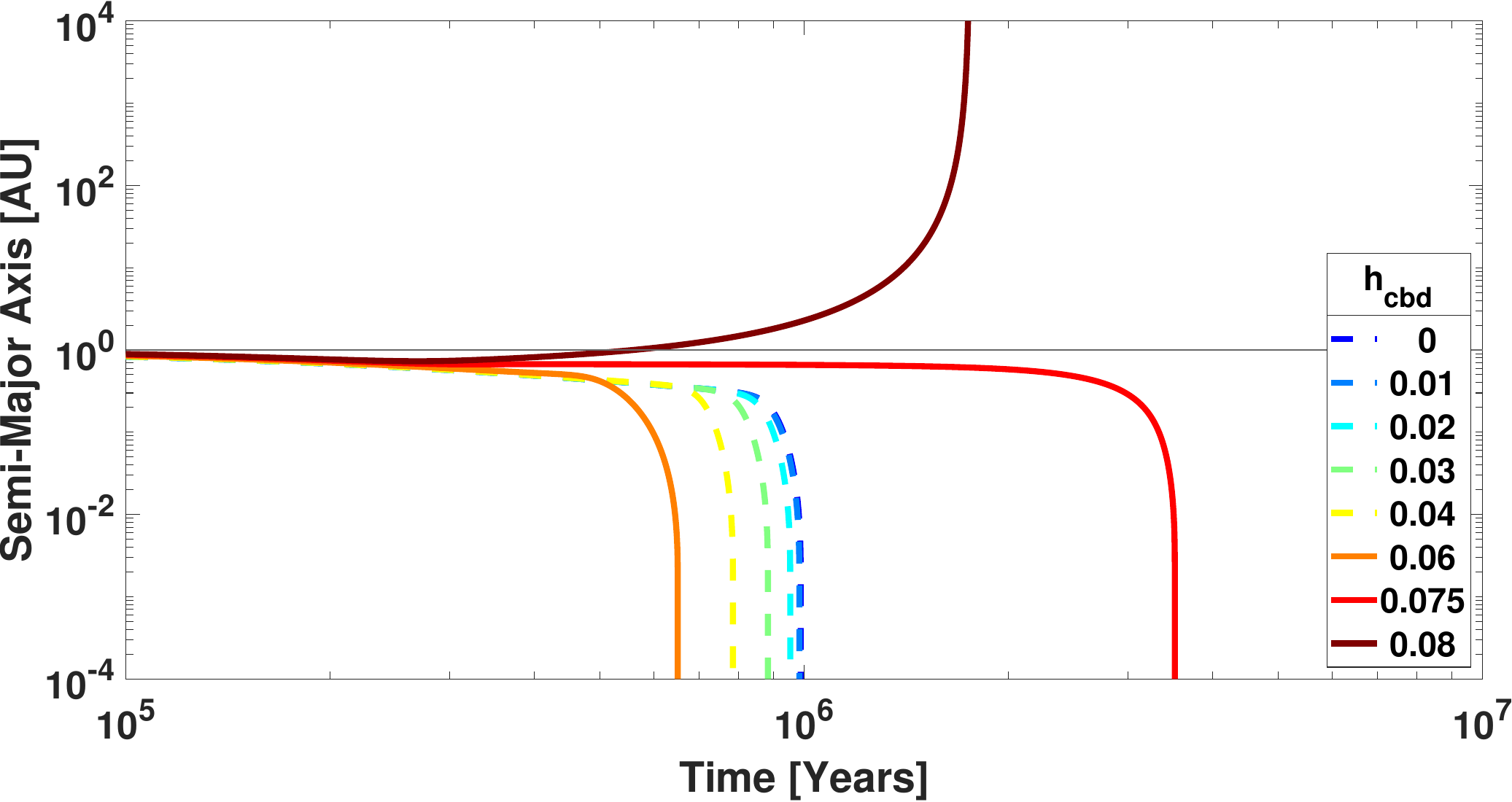}
    \caption{Semi-major axis as a function of time for variations in the disc aspect ratio $h_\mathrm{cbd}$ in the fiducial case ($p = 1/2$). Accreting binaries expand for $h_\mathrm{cbd} > h_\mathrm{crit}$, and contract for $h_\mathrm{cbd} < h_\mathrm{crit}$. The limiting case $h_\mathrm{cbd} = 0$ corresponds to a non-accreting binary. The dashed lines represent the orbits lying below the trend-reversing value ($h_\mathrm{cbd} < h_\mathrm{tr}$). } 
    \label{Plot_a_t_p05} 
\end{figure} 

In Fig. \ref{Plot_a_t_p05}, we show the temporal evolution of the semi-major axis in the fiducial case (corresponding to $p = 1/2$), for different values of the disc aspect ratio. We see that accreting BBHs contract for small disc aspect ratios, while they expand for large disc aspect ratios. In fact, there is a certain critical value of the disc aspect ratio, $h_\mathrm{crit} \sim 0.08$ (for fiducial parameters), above which accreting binaries undergo continuous orbital expansion and no coalescence is possible. Thus BBH mergers are prevented in thick discs, with aspect ratios exceeding a given critical value. On the other hand, all accreting binaries contract and merge in thin discs, provided that the aspect ratio is smaller than the critical value. This suggests the existence of a critical disc aspect ratio ($h_\mathrm{crit}$) separating contraction and expansion.

Moreover, we note that accreting binaries in thin discs merge faster with increasing disc aspect ratio, up to a certain transition value, $h_\mathrm{tr} \sim 0.06$ (for fiducial parameters). Beyond this point, the trend reverses and BBH mergers are slowed down for a further increase in disc thickness, until eventually reaching the critical value for expansion. 

\begin{figure}
  \centering
 \includegraphics[width=0.45\textwidth]{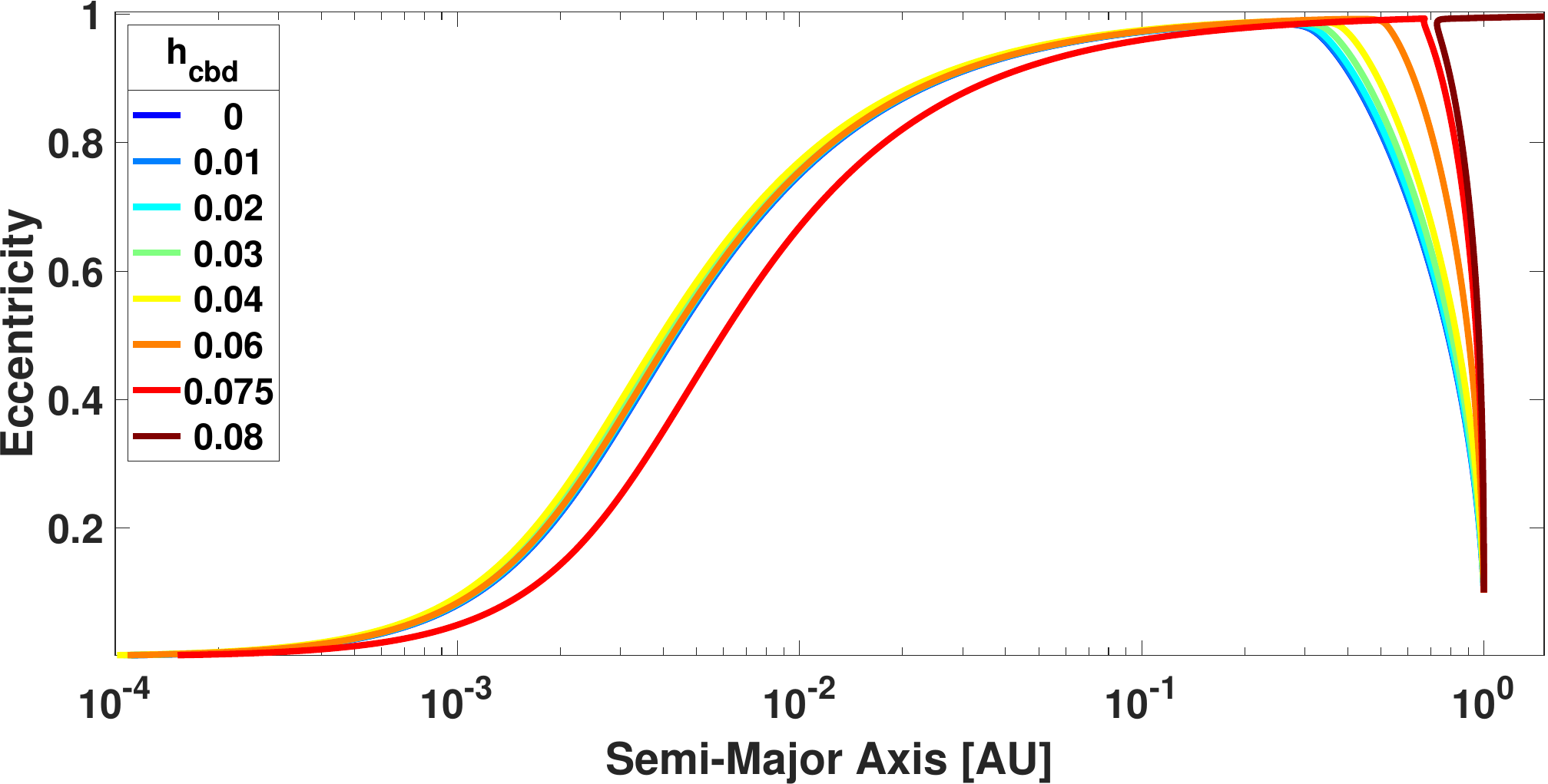}
    \caption{ Orbital eccentricity as a function of semi-major axis for different values of the disc aspect ratio $h_\mathrm{cbd}$. The eccentricity increases in the disc-driven regime (at large separations) and decreases in the GW-driven regime (at small separations). }
    \label{Plot_e_a_crit} 
\end{figure} 

In Fig. \ref{Plot_e_a_crit}, we plot the associated evolution of the orbital eccentricity as a function of the semi-major axis, for different values of the disc aspect ratio. For $h_\mathrm{cbd} < h_\mathrm{crit}$, we observe that the eccentricity strongly increases in the disc-dominated regime and then decreases in the GW-driven regime, while the binary separation continuously shrinks. An enhanced eccentricity growth in the disc-dominated regime can lead to a faster inspiral in the GW-driven regime, hence accelerated BBH mergers in AGN discs. 


\subsection{ Additional cases }

\begin{figure}
  \centering
 \includegraphics[width=0.45\textwidth]{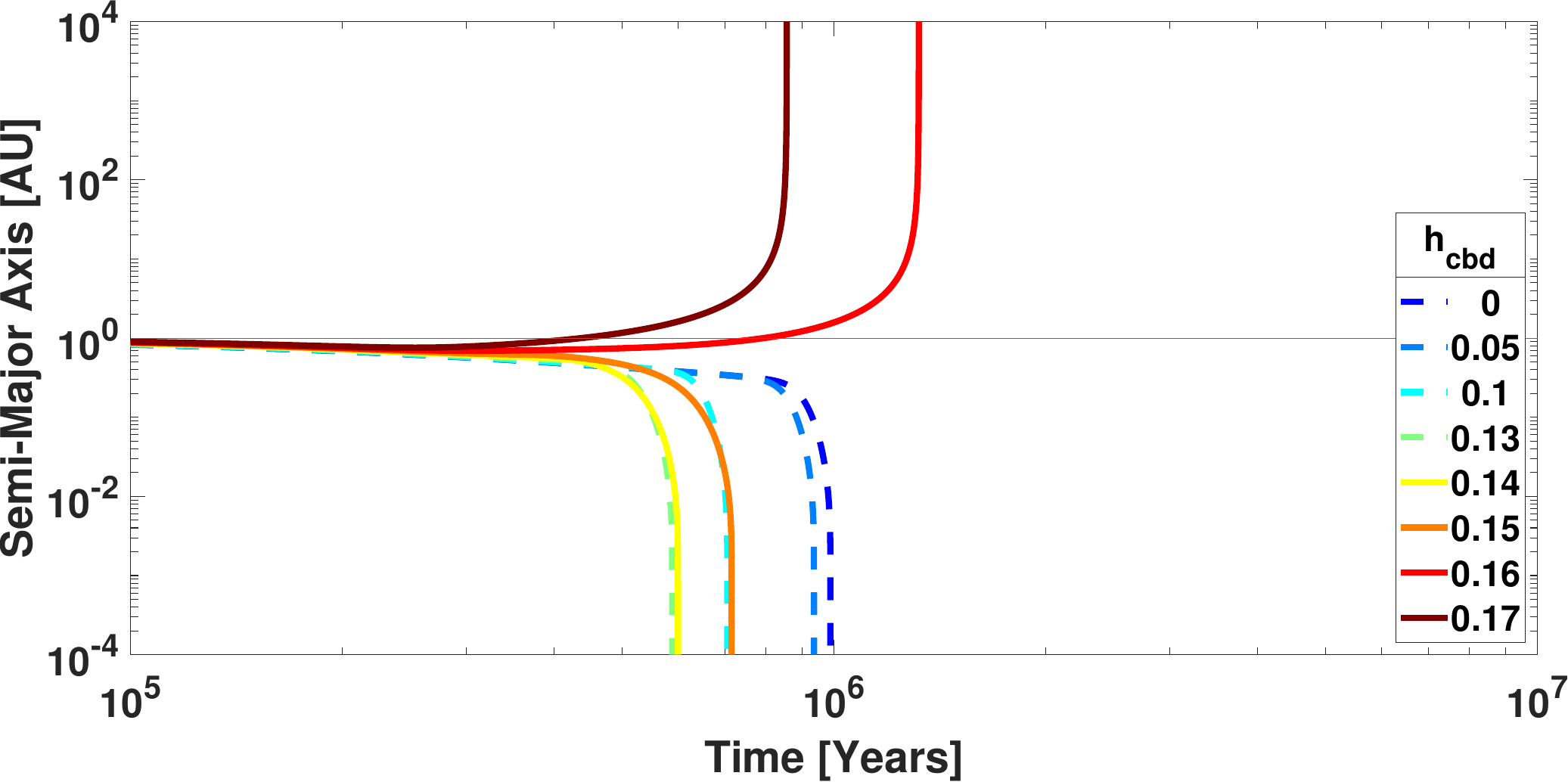}
    \caption{ Temporal evolution of the semi-major axis. Analogous to Fig. \ref{Plot_a_t_p05}, but in the case of $p = 1$. } 
    \label{Plot_a_t_p1} 
\end{figure} 

We next analyse radially declining accretion rate profiles with different power law slopes, $\dot{M}(r) \propto r^p$. Figure \ref{Plot_a_t_p1}  shows the temporal evolution of the semi-major axis for $p = 1$ (strong outflow). Compared to the fiducial case of $p = 1/2$ in Fig. \ref{Plot_a_t_p05}, the critical value of the disc aspect ratio is larger ($h_\mathrm{crit} \sim 0.16$), and accreting binaries can merge in thicker discs. Conversely, in the no-outflow case ($p = 0$), BBH merging can only occur for smaller disc aspect ratios ($h_\mathrm{crit} \sim 0.04$). Similar variations are also observed for the trend-reversal transition points ($h_\mathrm{tr}$). 
We summarise the key values of the disc aspect ratio corresponding to the three different power law slopes in Table \ref{Table_h_p}.  

\begin{table}
\caption{ Key values of the disc aspect ratio for three power law slopes characterising the accretion rate radial profile $\dot{M}(r) \propto r^p$.}
\centering
\begin{tabular}{c|c|c|c}
\hline
\hline
& $p=0$ & $p=1/2$ & $p=1$ \\
\hline
$h_\text{crit}$ & 0.04 & 0.075 & 0.16 \\ \hline
$h_\text{tr}$& 0.025 & 0.06 & 0.14 \\
\hline
\end{tabular} 
\label{Table_h_p}
\end{table}

\begin{figure}
  \centering
 \includegraphics[width=0.45\textwidth]{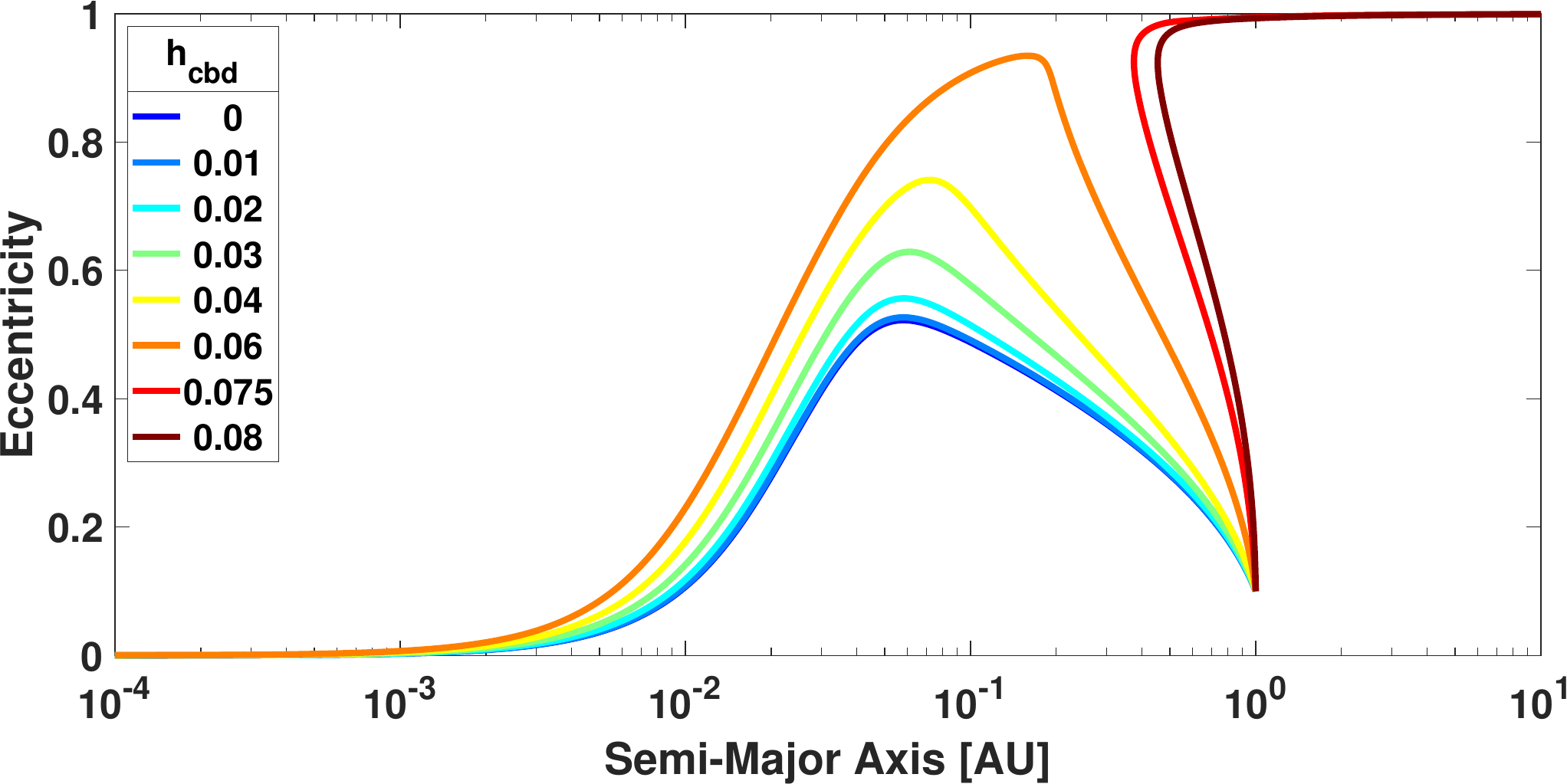}
    \caption{ Orbital eccentricity as a function of semi-major axis in the case of reduced eccentricity growth. Here the eccentricity growth terms in the disc-driven regime are reduced by a factor of ten with respect to the fiducial case (corresponding to $\sim 10 \%$ of the eccentricity growth represented in Fig. \ref{Plot_e_a_crit}). }
    \label{Plot_e_a_crit_reduced} 
\end{figure}

In addition, we also consider the case of reduced eccentricity growth. The increase in eccentricity is primarily due to the dominant outer Lindblad resonance operating in the disc-dominated regime (Sect. \ref{Section_model}). However, at high eccentricities, several competing higher-order resonances can be simultaneously present in the disc, and their opposite contributions to the eccentricity evolution ($e$-growth vs. $e$-damping) may partially cancel out. As a consequence, the eccentricity growth will be considerably reduced, and the maximal eccentricity reached will be significantly smaller than unity. An example of such a reduced eccentricity growth is shown in Fig. \ref{Plot_e_a_crit_reduced}. We observe that for small $h_\mathrm{cbd}$, the eccentricity increases up to $e \sim 0.5$ and subsequently decays as the binary undergoes GW-driven inspiral. Therefore, even in the case of reduced eccentricity growth, accreting binaries can still transition into the GW-dominated regime and eventually merge in thin discs. 

\section{Merger timescales of accreting binaries in AGN discs}
\label{Section_merger_timescale}

Following the disc-driven orbital decay and subsequent GW inspiral, accreting BBHs may eventually merge in thin discs (as long as $h_\mathrm{cbd}$ is smaller than the critical value). In our model framework, we can directly quantify the resulting BBH merger timescale, defined as the point where the numerical solution $a(t)$ or equivalently $e(t)$ approaches the abscissa axis. 

\begin{figure}
  \centering
 \includegraphics[width=0.45\textwidth]{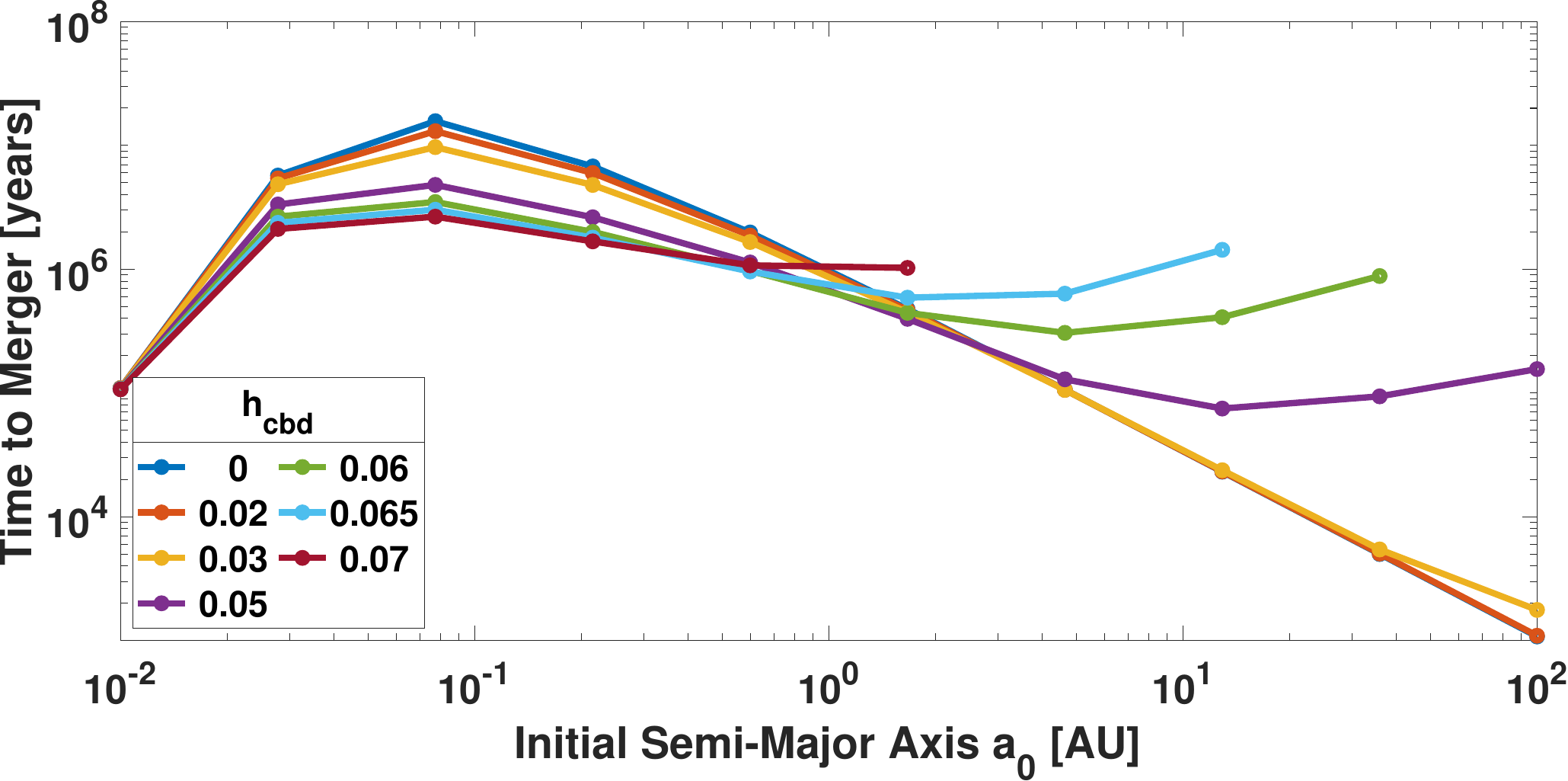}
    \caption{ Merger timescale ($\tau_\mathrm{merger}$) as a function of the initial semi-major axis ($a_0$) for different disc aspect ratios ($h_\mathrm{cbd}$). The merger time tends to decrease with increasing disc aspect ratio (for $h_\mathrm{cbd} < h_\mathrm{tr}$) at smaller $a_0$; whereas the merger time tends to increase with increasing disc aspect ratio at larger $a_0$. }
    \label{Plot_tau_a0_h} 
\end{figure}

Figure \ref{Plot_tau_a0_h} shows the merger time ($\tau_\mathrm{merger}$) of accreting binaries as a function of the initial semi-major axis $a_0$, for different values of the disc aspect ratio. We see that the BBH merger times mostly span the range $\tau_\mathrm{merger} \sim (10^5 - 10^7)$ years, depending on the initial separation, with a typical value of $\sim 10^6$ yr for $a_0 = 1$ AU. Such $\sim$Myr timescales are comparable to typical AGN disc lifetimes. We recall that the corresponding merger time would exceed the Hubble time ($\tau_\mathrm{GW} > 10^{10}$ yr) in the case of a purely GW-driven inspiral of a binary system located in the field \citep{Ishibashi_Groebner_2020}. Thus the merger timescale can be reduced by several orders of magnitude when including disc-binary interaction. 

It is interesting to note that the merger time of accreting BBHs can be significantly shorter than that of non-accreting counterparts (for $10^{-2} \lesssim a_0 \lesssim 10^0$ AU), with small $\tau_\mathrm{merger}$ values for moderate disc aspect ratios. Indeed, the merger time becomes shorter for larger disc aspect ratios, for a given initial separation (at $a_0 \sim 0.1$ AU). At an initial separation of $a_0 \sim 1$ AU, the merger time decreases with increasing disc aspect ratio, as long as $h_\mathrm{cbd} < h_\mathrm{tr}$; beyond this point, the merger time tends to increase with increasing $h_\mathrm{cbd}$, reflecting the trend reversal observed in Fig. \ref{Plot_a_t_p05}. 

At larger initial separations (beyond a few AU), the merger time increases with increasing disc aspect ratio, up to the critical value above which no merger occurs. On the other hand, the merger timescales become roughly independent of the disc aspect ratio, and converge towards $\tau_\mathrm{merger} \sim 10^5$ yr, for an initial separation of $a_0 \sim 0.01$ AU. This is likely the case of binaries in the GW-dominated regime, where disc-binary interaction is unimportant. We note that the $h_\mathrm{cbd}$-trend reversal behaviour is present on intermediate scales, while no trend reversal is observed at small ($a_0 < 0.1$ AU) or large ($a_0 > 10$ AU) separations. 

In addition, we verify that the merger timescale is always much shorter than the case of a purely GW-driven inspiral, even for reduced eccentricity growth. In fact, the typical merger time is still $\tau_\mathrm{merger} \sim 10^7$ years, for an eccentricity growth reduced by a factor of ten with respect to the fiducial case (cf. Sect. \ref{Section_contraction_expansion}). This suggests that the shortening of the merger timescale cannot be entirely attributed to the eccentricity growth and that accreting BBHs may still merge within the AGN disc lifetime --provided that the disc aspect ratio stays below the critical value.  

\subsection{ Dependence on binary parameters} 

\begin{figure}
  \centering
 \includegraphics[width=0.45\textwidth]{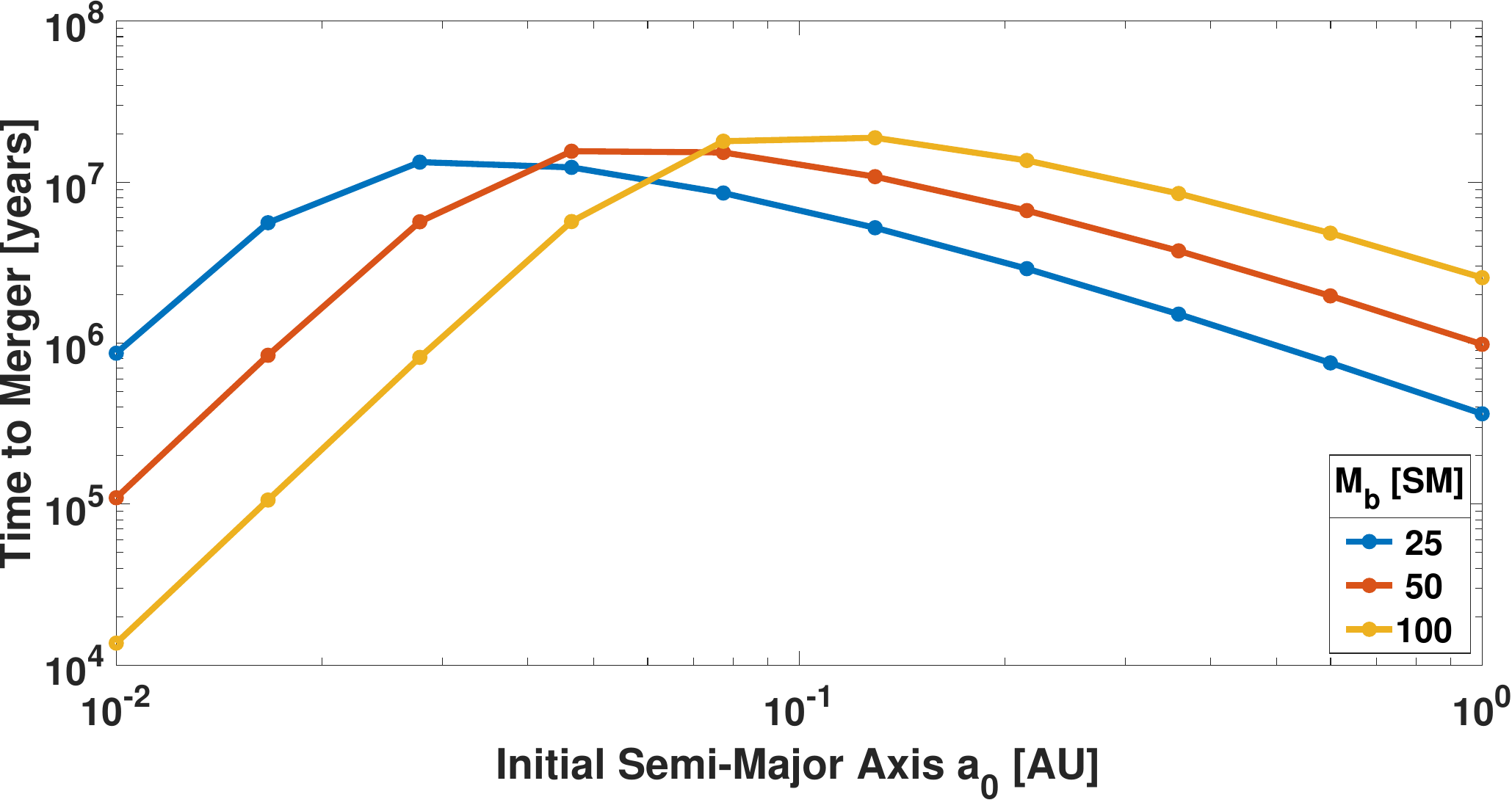}
    \caption{ Merger time versus initial semi-major axis for different binary mass $M_b$ (with fiducial parameters and $h_\mathrm{cbd} = 0.01$). }
    \label{Plot_tau_a0_Mb} 
\end{figure}

We next analyse the dependence of the merger timescale on the binary mass $M_b$ and mass ratio $q$. 
In Fig. \ref{Plot_tau_a0_Mb}, we plot the merger time as a function of the initial semi-major axis, for three different values of the total binary mass $M_b = (25, 50, 100) \, M_{\odot}$. At an initial separation of $a_0 = 1$ AU, the merger time is shorter for smaller binary mass; whereas at $a_0 = 0.01$ AU, the merger time is shorter for larger $M_b$. 
 
\begin{figure}
  \centering
 \includegraphics[width=0.45\textwidth]{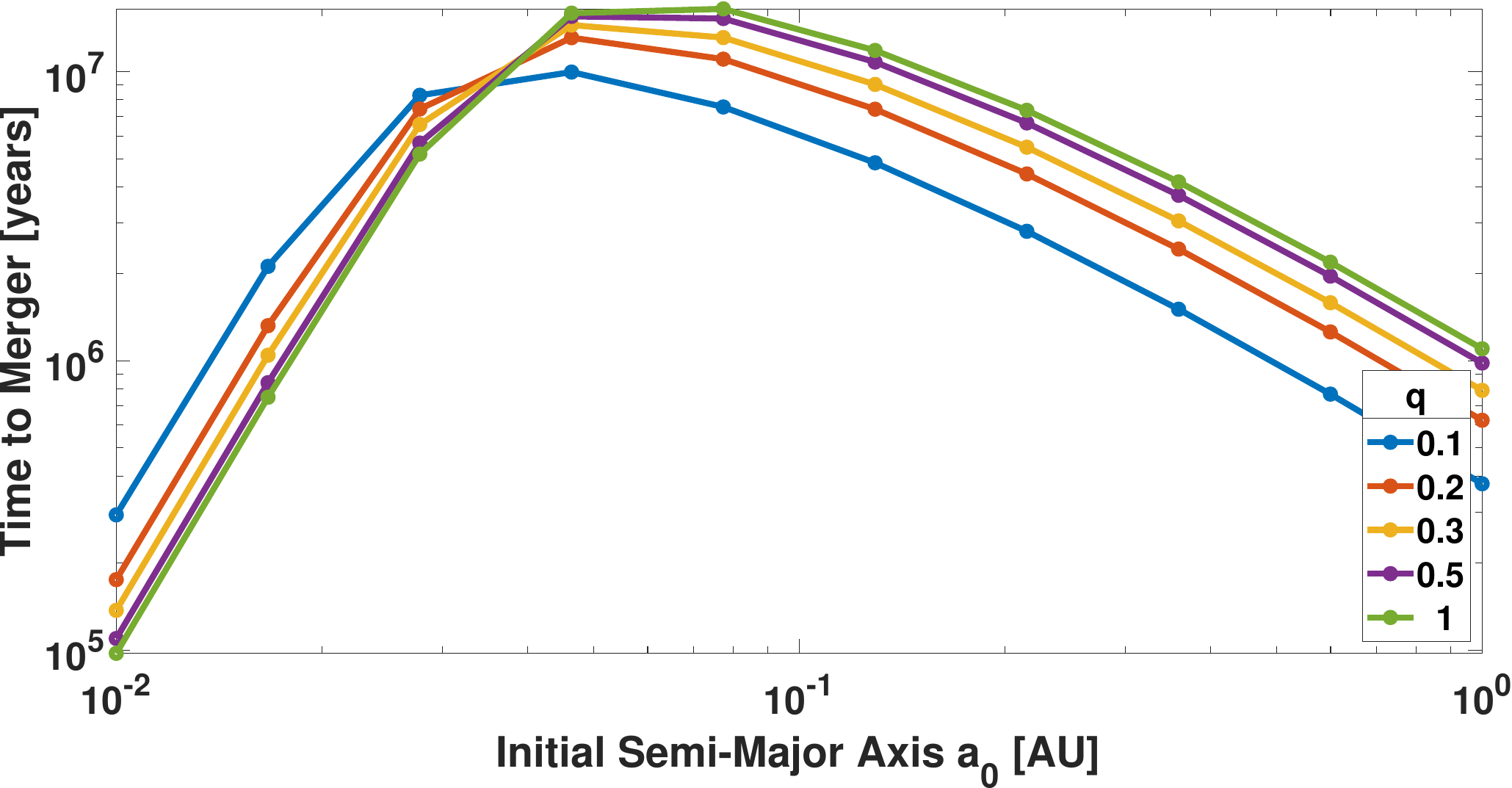}
    \caption{ Merger time versus initial semi-major axis for different mass ratio $q$ (with fiducial parameters and $h_\mathrm{cbd} = 0.01$). }
    \label{Plot_tau_a0_q} 
\end{figure}

Similarly, Fig. \ref{Plot_tau_a0_q} shows the merger time as a function of the initial semi-major axis, for five different values of the mass ratio $q = 0.1, 0.2, 0.3, 0.5, 1$. We see that at an initial separation of $a_0 = 1$ AU, the merger time is shorter for smaller mass ratios; whilst at $a_0 = 0.01$ AU, the merger time is shorter for larger $q$. 

Combining the above two trends, we obtain that smaller binary mass and smaller mass ratio lead to shorter merger timescales at large initial separations ($a_0 \sim 1$ AU); in contrast, shorter merger times are obtained for larger binary mass and larger mass ratio at small initial separations ($a_0 \sim 0.01$ AU). The two opposite trends may be interpreted in terms of the different analytic scalings in the disc-driven and GW-driven regimes. Given that we are dealing with merging binaries, the negative viscous term should ultimately dominate the disc-driven evolution equation. The dependence of the contraction rate on binary mass and mass ratio in the two regimes are then given by 
\begin{equation}
\dot{a}_\mathrm{disc} \propto - \frac{(1+q)^2}{q} M_b^{-3/2}
\end{equation}
\begin{equation}
\dot{a}_\mathrm{GW} \propto - \frac{q}{(1+q)^2} M_b^{3}
\end{equation}  
Thus smaller $M_b$ and smaller $q$ lead to greater $\dot{a}_\mathrm{disc}$; conversely, larger $M_b$ and larger $q$ lead to greater $\dot{a}_\mathrm{GW}$. An enhanced shrinkage rate (greater $\dot{a}$) then naturally lead to shorter merger timescales. Overall, a combination of moderate disc aspect ratio, smaller binary mass, and smaller mass ratio can lead to shorter merger timescales for binaries initialised at $a_0 \sim 1$ AU. 

\section{GW merger rates of accreting binaries in AGN discs}
\label{Section_GW_rates}

A rough estimate of the resulting GW merger rate density can be obtained from the merger timescales derived in the previous section. The BBH merger rate in AGN discs is parametrised by \citep[e.g.][]{McKernan_et_2018, Ford_McKernan_2022}
\begin{equation}
\mathcal{R} = \frac{n_\text{AGN} N_\text{BH} f_d f_b}{\tau_\mathrm{merger}},
\label{GW_rate}
\end{equation}
where $n_\text{AGN}$ is the number density of AGNs (per $\mathrm{Gpc^{-3}}$), $N_\text{BH}$ is the number of stellar-mass BHs in the galactic nucleus (within the central $\sim\text{pc}^3$), $f_d$ is the fraction of BHs in the AGN disc, $f_b$ is the fraction of such BHs residing in binaries, and $\tau_\mathrm{merger}$ is the merger time. 
The following values are taken as fiducial parameters: $n_\text{AGN} = 5 \times 10^4 \, \text{Gpc}^{-3}$, $N_\text{BH} = 10^4$, $f_d = 0.01$, and $f_b = 0.1$ \citep[cf.][for more details]{Groebner_et_2020}. 

We recall that in most previous works, the merger timescale is not explicitly calculated and binaries are simply assumed to merge within the AGN disc lifetime \citep{McKernan_et_2018, Tagawa_et_2020}. We instead compute the actual merger times of accreting BBHs in AGN discs, as a function of the different binary parameters and disc configurations (see Section \ref{Section_merger_timescale}). Based on equation \ref{GW_rate}, we derive associated BBH merger rates in AGN discs in the range $\mathcal{R} \sim (0.2 - 5) \, \text{Gpc}^{-3} \text{yr}^{-1}$, depending on the different combinations of the underlying binary/disc parameters. For a typical BBH merger time of a $\sim$Myr, the corresponding GW merger rate is about $\mathcal{R} \sim 0.5 \, \text{Gpc}^{-3} \text{yr}^{-1}$. 

These values are consistent with the BBH merger rate of $\mathcal{R} \sim 1.2 \, \text{Gpc}^{-3} \text{yr}^{-1}$ estimated in the inner regions of AGN discs, as well as the merger rate of $\mathcal{R} \sim 3 \, \text{Gpc}^{-3} \text{yr}^{-1}$ expected in AGN self-gravitating discs \citep{Bartos_et_2017, Stone_et_2017}. By modelling the formation and evolution of binaries in AGN discs (including several physical processes, such as dynamical friction and type I/II migration), \citet{Tagawa_et_2020} obtain a BBH merger rate in the range $\mathcal{R} \sim (0.02 - 60) \, \text{Gpc}^{-3} \text{yr}^{-1}$. More recently, \citet{Ford_McKernan_2022} argue that the AGN disc channel may account for up to $\sim (25-80) \%$ of the observationally inferred BBH merger rate. 

The current LVK GWTC-3 measurements indicate BBH merger rates in the range $\mathcal{R} \sim (17.9 - 44) \, \text{Gpc}^{-3} \text{yr}^{-1}$, with a characteristic value of $\mathcal{R} \sim 23.9 \, \text{Gpc}^{-3} \text{yr}^{-1}$ \citep{LVK_GWTC_3_pop}. By comparison, our predicted BBH merger rate in AGN discs may provide a non-negligible contribution to the observed rate, possibly of the order of a few to tens of percent. In fact, our derived range of $\mathcal{R} \sim (0.2 - 5) \, \text{Gpc}^{-3} \text{yr}^{-1}$ may be considered as a conservative estimate of the actual BBH merger rate in AGN discs, as the fiducial model values lie on the lower end of the astrophysically plausible range (e.g. the chosen value of $n_\text{AGN}$ is appropriate for rare luminous quasar populations).

Since the BBH merger rate scales inversely with the merger timescale ($\mathcal{R} \propto 1/\tau_\mathrm{merger}$), we can qualitatively understand the physical trends from the $\tau_\mathrm{merger}$-dependence (Section \ref{Section_merger_timescale}). For instance, higher BBH merger rates can be expected for less massive binaries with smaller mass ratio evolving in discs with moderate aspect ratio. In particular, a favourable combination of the underlying parameters (moderate $h_\mathrm{cbd}$, smaller $M_b$, and smaller $q$, for an initial separation of $a_0 \sim 1$ AU) may give rise to the highest merger rates. Interestingly, accreting BBHs in AGN discs can experience accelerated mergers that lead to shorter merger times, hence higher GW event rates, compared to non-accreting counterparts. This implies that accreting BBHs in AGN discs could provide a significant contribution to the observed GW merger rate.  


\section{ Discussion }
\label{Section_discussion}

Contrasting results have been reported in the literature concerning the orbital evolution of binaries in circumbinary discs. In principle, the interaction between the binary and the surrounding gaseous disc can lead to either orbital contraction or expansion. While the classic theory of disc-binary interaction commonly suggests contraction \citep[][]{Pringle_1991, Artymowicz_et_1991, Lubow_Artymowicz_2000}, recent numerical simulations indicate that accreting binaries tend to expand over time \citep{Munoz_et_2019, Moody_et_2019, Duffell_et_2020}. 

2D hydrodynamic simulations of circumbinary accretion indicate that the binary can actually gain net angular momentum, implying a positive $\langle \dot{a} \rangle > 0$ \citep{Munoz_et_2019}. 3D hydrodynamic simulations show that the total torque is positive, such that the binary separation increases with time, for both aligned and misaligned discs \citep{Moody_et_2019}. Similarly, \citet{Duffell_et_2020} obtain that the net torque on the binary is (almost) always positive, with little dependence on mass ratio and disc viscosity. We note that most of these numerical simulations consider fixed binary orbits. Moreover, a large disc aspect ratio (of order $h \sim 0.1$) is usually assumed in the simulations, due to computational limitations. However, in more realistic situations, the accretion discs in AGNs are likely thinner ($h \sim 10^{-2} - 10^{-3}$), and the binary orbit must evolve as a result of the interactions with the gaseous disc. 

More recently, the conclusion that accreting binaries undergo expansion rather than contraction has been challenged by other studies. In particular, it has been argued that the direction of the binary orbital evolution can depend on the disc aspect ratio: thicker discs lead to binary expansion, while thinner discs lead to orbital contraction. For instance, 2D isothermal hydrodynamic simulations of accreting binaries indicate a transition from outspiral in thick discs to inspiral in thin discs, with a critical disc aspect ratio of $h \sim 0.04$ \citep{Tiede_et_2020}. The importance of the disc aspect ratio is emphasised by \citet{Heath_Nixon_2020}, who perform 3D simulations of `live' binaries --as opposed to fixed binary orbits assumed in previous simulations-- showing that accreting binaries are prone to shrink in realistic thin discs. They suggest that the critical disc aspect ratio separating orbital expansion and contraction is larger than the value reported in \citet{Tiede_et_2020}, possibly around $h \sim 0.1$, with the actual dividing line being parameter-dependent. 

In this context, we analytically investigate the evolution of accreting binaries in AGN discs by considering the torque balance based on angular momentum conservation. A novel feature of our analysis is the coupled `disc+GW'-driven evolution: we follow the temporal evolution of the binary orbital parameters through the transition from the disc-driven regime at large separations into the GW-driven regime at small separations, i.e. from the initial disc-binary interactions to possible GW-driven coalescence. In broad agreement with other studies, we find that accreting binaries shrink in thin discs, while they tend to expand in thick discs. More precisely, we obtain binary contraction for disc aspect ratios smaller than a certain critical value ($h_\mathrm{crit}$), above which accreting binaries undergo orbital expansion. We derive critical disc aspect ratios in the range $h_\mathrm{crit} \sim 0.04 - 0.16$, with $h_\mathrm{crit} \sim 0.08$ for fiducial parameters, which are consistent with the values obtained in previous numerical simulations \citep{Tiede_et_2020, Heath_Nixon_2020}.

While GW emission always tends to shrink the binary orbit, accreting BBHs can either contract or expand in the disc-dominated regime (depending on the disc aspect ratio). In the particular case of circular binaries, the binary orbital evolution is just described by the rate of change in the semi-major axis (equation \ref{Eq_a_disc}). In this simple case, the positive accretion terms tend to counter-act the negative viscous term, so that the inclusion of accretion always tends to reduce the shrinkage rate, possibly leading to orbital expansion. This may be consistent with simulation results indicating that accreting circular binaries gain net positive angular momentum and undergo expansion in most cases \citep{Moody_et_2019}. 

However, in the more general case of eccentric binaries, the BBH orbital evolution is likely modified due to the complex coupling with the orbital eccentricity. Although accreting BBHs may gain angular momentum from the disc, they can still shrink and merge in thin discs, provided that the disc aspect ratio is smaller than $h_\mathrm{crit}$. Actually, there are hints of a reduced expansion rate for an eccentric binary compared to a circular counterpart in hydrodynamical simulations \citep{Munoz_et_2019}, suggesting that the eccentricity could play an important role in the orbital evolution of accreting binaries \citep{Duffell_et_2020}.  

In our picture, accreting binaries contract and eventually merge in thin discs, and their mergers can be accelerated for moderate disc aspect ratios (provided that $< h_\mathrm{crit}$). Indeed, accreting BBHs merge faster with increasing disc aspect ratio, up to a certain transition point ($h_\mathrm{tr}$), above which the trend reverses (for intermediate $a_0$). This non-monotonic dependence on the disc aspect ratio is likely due to a combination of eccentricity growth in the disc-dominated regime and enhanced orbital decay in the GW-driven regime. 

The increase in eccentricity is mostly driven by the outer Lindblad resonance, which dominates the eccentricity evolution in our fiducial case. We caution that in other situations (e.g. high eccentricities and/or thick discs), the effect of eccentricity excitation may be significantly reduced and become inefficient, due to the action of competing resonances simultaneously present in the disc. For instance, several overlapping resonances may operate at high eccentricities, such that the effect of eccentricity growth will be partially cancelled out as a result of eccentricity damping by corotation and inner Lindblad resonances. Nevertheless, even if the eccentricity growth is reduced in the disc-dominated regime, some accreting binaries may still be able to transition into the GW-driven regime and ultimately merge in thin discs (Sect. \ref{Section_contraction_expansion}). A more detailed study of the eccentricity evolution ($e$-growth vs. $e$-damping) in discs of varying thickness will be required in future work. 
 
The global trends with disc aspect ratio ($h_\mathrm{crit}, h_\mathrm{tr}$) are directly imprinted on the corresponding BBH merger timescales and associated GW merger rates. Contrary to common expectations, accreting binaries can experience accelerated mergers and have shorter merger times compared to non-accreting counterparts. The merger time decreases with increasing disc aspect ratio up to a certain transition point, such that shorter $\tau_\mathrm{merger}$ are obtained for moderate disc aspect ratios (at intermediate $a_0$). Typical merger times are on the order of $\sim$ Myr, comparable to the lifetimes of AGN discs. In addition, the merger timescale also depends on BBH parameters, such as binary mass and mass ratio. The resulting GW merger rates are of the order of $\sim 1 \, \text{Gpc}^{-3} \text{yr}^{-1}$, which should be viewed as a conservative estimate. Therefore accreting BBHs in AGN discs may provide a non-negligible contribution to the GW merger rate observed by the LVK collaboration.  
 

\section{ Conclusion }
\label{Section_conclusion}

We consider the evolution of accreting BBHs in AGN discs as a potential formation path for GW sources. In our analytic framework, we obtain a critical disc aspect ratio ($h_\mathrm{crit}$) separating orbital contraction and expansion: accreting binaries expand in thick discs, while they contract and merge in thin discs. In contrast to common belief, we find that some accreting BBHs can undergo accelerated mergers with respect to non-accreting counterparts. 

A novel aspect of our work is that we follow the coupled `disc+GW'-driven evolution, which allow us to quantify the BBH merger timescale in AGN discs and the associated GW merger rate. We predict typical BBH merger timescales of $\tau_\mathrm{merger} \sim (10^5 - 10^7)$ yr, and corresponding GW merger rates in the range $\mathcal{R} \sim (0.2 - 5) \, \text{Gpc}^{-3} \text{yr}^{-1}$, for this particular BBH-in-AGN channel. 

Overall, the global picture of BBH evolution in AGN discs seems to be more complex than previously thought. There is no simple one-to-one relation whereby mass accretion always leads to binary expansion. Instead, accreting binaries in AGN discs can either expand or contract depending on the complex interplay between disc-binary interaction and subsequent GW emission. The final outcome may be strongly parameter-dependent, and a wider parameter space study -- including the subtle coupling between orbital eccentricity and disc aspect ratio -- needs to be investigated in future numerical simulations.


\section*{Data availability}

No new data were generated or analysed in support of this research.


\bibliographystyle{mn2e}
\bibliography{biblio.bib}

\label{lastpage}

\end{document}